\documentclass[a4paper,11pt]{article}
\usepackage{xcolor}
\usepackage{graphicx}
\usepackage{wrapfig}
\usepackage{subcaption}
\usepackage{float}
\usepackage[export]{adjustbox}
\usepackage[font=small,skip=10pt]{caption}
\usepackage{lineno}
\usepackage{amsmath, amssymb, amsfonts,indentfirst,graphicx,color,anysize}
\marginsize{3cm}{3cm}{2cm}{2cm}
\title{
	\includegraphics[width=0.35\textwidth]{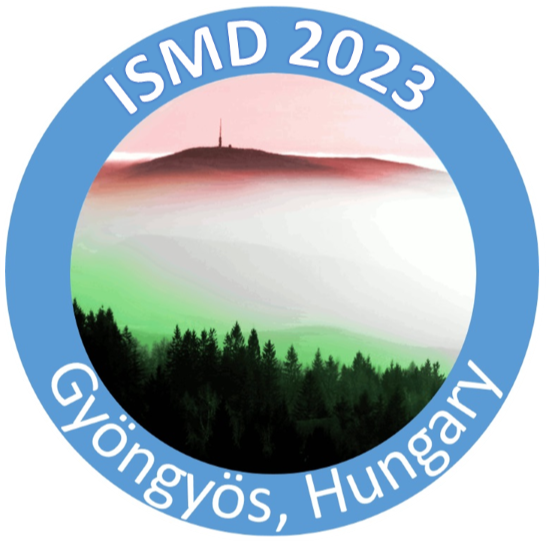}\\[1cm]
	\textbf{First-order event plane correlated directed and triangular flow from fixed-target energies at RHIC-STAR}}
\author{{Sharang Rav Sharma (for STAR Collaboration)}\\[1ex]
	Indian Institute of Science Education and Research (IISER) Tirupati\\
}
\begin{document}

\maketitle
\begin{abstract} 
We report the measurement of first-order event plane correlated directed flow $(v_1)$ and triangular flow $v_3$ for identified hadrons ($\pi^{\pm}$, $K^{\pm}$, and $p$), net-particle (net-K, net-p), and light nuclei ($d$ and $t$) in Au+Au collisions at $\sqrt{s_{\text{NN}}}$ = 3.2, 3.5, and 3.9 GeV in fixed-target mode from the second phase of beam energy scan (BES-II) program at RHIC-STAR. The $v_1$ slopes at mid-rapidity for identified hadrons and net-particles except $\pi^{+}$ are found to be positive, implying the effect of dominant repulsive baryonic interactions. The slope of $v_1$ for net-kaon undergoes a sign change from negative to positive at a lower collision energy compared to net-proton. An approximate atomic mass number scaling is observed in the measured $v_1$ slopes of light nuclei at mid-rapidity, which favours the nucleon coalescence mechanism for the production of light nuclei. The $v_3$ slope for all particles decreases in magnitude with increasing collision energy, suggesting a notable integrated impact of the mean-field, baryon stopping, and collision geometry at lower collision energies.
\end{abstract}

\section{Introduction}
The primary objective of ultra-relativistic heavy-ion collisions at the Relativistic Heavy Ion Collider (RHIC) and the Large Hadron Collider (LHC) is to create and characterize a novel state of matter with partonic degrees of freedom, known as the Quark-Gluon Plasma (QGP). This state of strongly interacting matter is hypothesized to have been present during the initial microseconds following the Big Bang, and gaining an understanding of its properties holds the potential to offer insights into the evolution of the universe \cite{Ref1}.
The lattice Quantum Chromodynamics (QCD) predicts a crossover region between the hadron gas and Quark-Gluon Plasma (QGP) around T=154 MeV at low $\mu_B$ \cite{Ref3}. At lower temperatures and higher $\mu_B$, QCD-based models suggest a first-order phase transition concluding at a conjectured QCD critical point \cite{Ref4}. Numerous experimental observables measured at RHIC and LHC have presented compelling evidence of Quark-Gluon Plasma (QGP) formation for matter near $\mu_B = 0$. However, experimental confirmation of the existence of a critical point and a first-order phase transition at higher $\mu_B$ is still pending.
Many signatures of QGP formation and associated characteristics of the medium have been proposed. This proceeding will briefly explore one of the proposed signatures, namely, anisotropic flow. The patterns of azimuthal anisotropy in particle production are commonly referred to as flow. In heavy-ion collisions, it can be obtained by studying the Fourier expansion of the azimuthal angle $(\phi)$ distribution of produced particles with respect to the event plane angle $(\Psi_{n})$. \\
The particle azimuthal angle distribution is written in the form of a Fourier series \cite{Ref4},

\begin{equation}
    E\frac{\textrm{d}^3N}{\textrm{d}p^3} = \frac{\textrm{d}^2N}{2\pi p_T 
    \textrm{d}p_T \textrm{d}y} \, \left\{1 \,+\, \sum_{n \geq 1} 2 \, v_n 
    \cos\left[ n \left(\phi \,-\, \Psi_n \right) \right] \right\},
\end{equation}

\noindent 
where $p_T$, $y$, $\phi$, and $\Psi_n$ are particle transverse momentum, rapidity, azimuthal angle of the particle and the $n^{th}$ order event plane angle, respectively. 
The various (order n) coefficients in this expansion are defined as:
    \begin{equation}
        v_n = \langle \cos[n(\phi - \Psi_R)]\rangle
    \end{equation}
    
 \noindent
The angular brackets in the definition denote an average over many particles and events \cite{Ref4}. These flow anisotropy parameters offer an insight into collective hydrodynamic expansion and transport properties of the produced medium at higher collision energies, while they are sensitive to the compressibility of the nuclear matter and nuclear equation of state at lower collision energies. The first three Fourier expansion coefficients $v_1$ (directed flow), $v_2$ (elliptic flow) and $v_3$ (triangular flow) are sensitive probes for studying the properties of the matter created in high-energy nuclear collisions.\\
At higher energies (nucleon-nucleon center-of-mass energy $\sqrt{s_{NN}} \gtrsim$ 27 GeV), where the transit time of colliding nuclei $2R/\gamma\beta$ is smaller than the typical production time of particles~\cite{Ref5}, flow harmonics are predominantly influenced by the collective expansion of the initial partonic density distribution~\cite{Ref6}. Conversely, at lower energies, the shadowing effect caused by passing spectator nucleons becomes significant. For $\sqrt{s_{NN}} \lesssim $ 4 GeV, nuclear mean-field effects contribute to the observed azimuthal anisotropies~\cite{Ref7}. Numerous studies indicate that flow coefficients are notably sensitive to the incompressibility of nuclear matter $(\kappa)$ in the high baryon density region~\cite{Ref8}. Comparing experimental data with results from theoretical transport models can provide constraints on $\kappa$, offering valuable insights into nuclear Equation of State (EOS).\\
\noindent
The $v_1$, sensitive to early collision dynamics, is proposed as a signature of first-order phase transition based on a hydrodynamic calculation. These calculations, whose equation of state incorporates a first-order phase transition from hadronic matter to QGP, predict a non-monotonic variation of the slope of the directed flow of baryons (and net-baryons) around midrapidity as a function of beam energy~\cite{Ref9}.\\
\noindent
The tradional $v_3$, third order flow coefficient typically results from fluctuations in shape of the initial condition and is not correlated to the reaction plane. Contrary to this, an observation has been made by the HADES collaboration at $\sqrt{s_{\text{NN}}}$ = 2.4 GeV Au+Au collisions regarding a noticeable triangular flow, correlated with the first order event plane ($\Psi_{1}$) \cite{Ref6}. $v_3$ is also observed to be sensitive to the equation of state and can serve as a new tool to explore the time dependence of the pressure during the heavy-ion collision~\cite{Ref10}. The evolution of $v_3$ is influenced by two crucial factors: the first involves the appropriate geometry determined by stopping, the passing time of spectators, and the expansion of the fireball; the second entails a potential within the responsive medium that propels the collective motion of particles.

\section{Dataset and Event Selection Cuts}
In these proceedings, we present the results of first order event plane $(\Psi_1)$ correlated $v_1$ and $v_3$ for identified hadrons ($\pi^{\pm}$, $K^{\pm}$, and $p$), net-particle (net-K, net-p), and light nuclei ($d$ and $t$) in Au+Au collisions at $\sqrt{s_{NN}}$ = 3.2, 3.5, and 3.9 GeV using the fixed target (FXT) data from the Solenoidal Tracker at RHIC (STAR) experiment. The FXT setup was implemented at STAR to explore the region of high baryon chemical potential ($\mu_{B}$) on the QCD phase diagram. This data was collected during the second phase of the Beam Energy Scan program (BES-II) (2019-2020) after incorporating various detector upgrades.\\
\\
In FXT mode, where the target is stationary at one end of the TPC, we apply a vertex cut along the z-direction ($v_z$) within [198, 202] cm. For the x and y directions, we set the $V_r$ ($\sqrt{V_x^2 +V_y^2}$) less than 2 cm centered around (0, -2).

\section{Analysis Details}
\subsection{Track Quality Cuts}
\noindent To ensure the quality of primary tracks, tracks with the transverse momentum $p_{T}$ $<$ 0.2 GeV/c are excluded. Additionally, we mandate the utilization of a minimum of 15 fit points and 52$\%$ of the total possible fit points in the track fitting process. The selection criterion involves choosing dE/dx hit points $\ge$ 10. Furthermore, the distance of closest approach (DCA) is set to $<$ 3 cm.
\subsection{Particle Identification}
\noindent The identification of charged particles in STAR is done by the combination of Time Projection Chamber (TPC) and Time of Flight (TOF) detectors. For low-momentum particles, TPC is used, whereas for particles with intermediate or high momenta ($p_T>$1 GeV/c), the TOF is used. TPC uses the ionization energy loss (dE/dx) of the charged particles passing through it for particle identification. Using dE/dx information, the z variable is defined:
\begin{equation}
    z = ln(\frac{\langle dE/dx \rangle}{\langle dE/dx\rangle_{X}^{B}}),
\end{equation}

\noindent
where $\langle dE/dX \rangle_X^{B}$ is the expected energy loss based on the Bichsel function and X is the particle type \cite{Ref11}. The raw yield from the TOF are obtained using the variable mass square ($m^2$), given by 
\begin{equation}
    m^2 = p^2 \left(\frac{c^2 T^2}{L^2} - 1 \right), 
\end{equation}
\noindent
where, p, T , L, and c are the momentum, time of travel by the particle, path length and speed of light, respectively. The left panel of Fig 1 shows the average dE/dx of measured charged particles plotted as a function of “rigidity” (i.e., momentum/charge) of the particles. The curves represent the Bichsel expectation values. The right panel of Fig 1 shows the inverse of particle velocity in unit of the speed of light $1/\beta$, as a function of rigidity. The expected values of $1/\beta$ for pions, kaons, and protons are shown as the curves.

\begin{figure}[H]
\centering
\begin{subfigure}{0.49\textwidth}
\includegraphics[scale = 0.225]{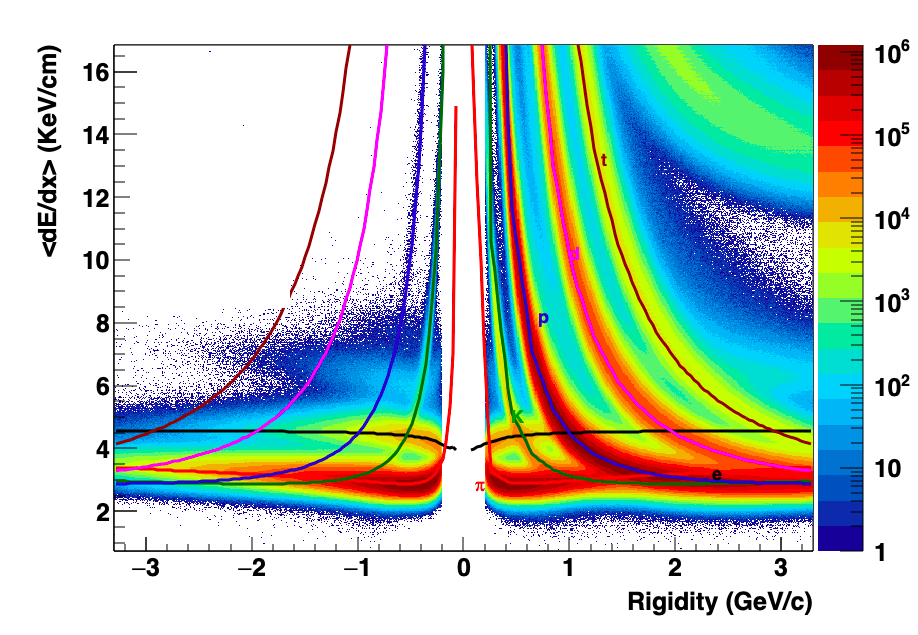}
\end{subfigure} 
    \begin{subfigure}{0.49\textwidth}
        \includegraphics[scale = 0.2575]{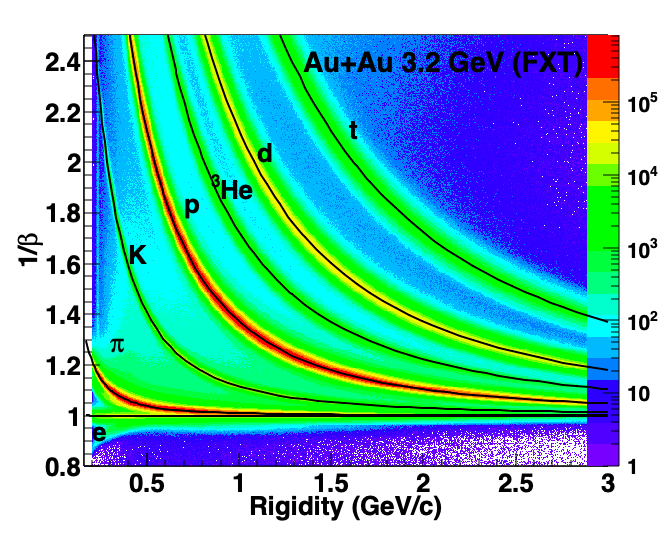}
    \end{subfigure}
    \caption{($\langle dE/dx \rangle$ (left panel) and 1/$\beta$ from TOF (right panel) of charged particles in Au+Au collisions at $\sqrt{s_{\text{NN}}}$ = 3.2 GeV.}
\end{figure}
In this analysis, for the identification of pion, kaon, and proton, we require TPC $n\sigma$ (z/R; R: TPC resolution)and TOF $m^{2}$ cuts which are listed in the Table 1. In addition to $m^{2}$ cut a momentum dependent z cut is implemented for light nuclei identification. 

\begin{center}
 \captionof{table}{Particle identification cuts}
    \begin{tabular}{|c|c|}
        \hline
         pion  &  $|n\sigma_{\pi}| <$ 3 and $-0.1< m^2 <0.15$ ($(GeV/c^2)^2$)        \\ \hline
         kaon &  $|n\sigma_{K}| <$ 3 and $0.16< m^2 <0.36$ ($(GeV/c^2)^2$)           \\ \hline
         proton &  $|n\sigma_{p}| <$ 2 and $-0.6< m^2 <1.2$ ($(GeV/c^2)^2$)          \\ \hline
    deuteron &   momentum dependent z cut and $3.15< m^2 <3.88$ ($(GeV/c^2)^2$)          \\ \hline
    triton &   momentum dependent z cut and $7.01< m^2 <8.75$ ($(GeV/c^2)^2$)          \\ \hline
         
    \end{tabular}
\end{center}

\noindent
\subsection{Event Plane Reconstruction}
%The flow coefficients are extracted from the azimuthal angle of particles relative to the azimuth of the reaction plane. However, the reaction plane angle cannot be directly measured in experiments; instead,
The event plane angle can be estimated from the particle azimuthal distribution on an event-by-event basis.
In our calculations, we utilized the first-order event plane angle $\Psi_{1}$, which is measured using the Event Plane Detector (EPD).
In order to calculate the first-order event plane angle, firstly, we construct the Q vector from particle’s azimuthal angle.
\begin{equation}
    \vec{Q} = (Q_{x},Q_{y})  = \left(\sum_{i}w_i\cos(\phi_i), \sum_{i}
    w_i \sin(\phi_i) \right),
\end{equation}
\begin{equation}
    \psi_1 = \tan^{-1}(Q_y/Q_x),
\end{equation}

\noindent
where sum extends over all detected hits $i$, and $\phi_i$ is the azimuthal angle in the laboratory frame, and $w_i$ is the weight for the $i^{th}$ hits, here we use the nMip as the weight, which is the calibrated ADC value. The $\psi_1$ is the first-order event plane angle. The first order event plane evaluated from spectator flow at forward rapidities is close to the reaction plane, defined as the plane subtended by the impact parameter vector and the beam axis~\cite{Ref4}. In order to mitigate acceptance correlations arising from the imperfect detector, it is essential to render the event plane angle distribution isotropic or flat. Consequently, a procedure for flattening the event plane angle distribution becomes necessary. In this analysis, we have implemented re-centering and shift corrections to extract a flat event plane angle distribution~\cite{Ref4}.\\
\subsection{Event Plane Resolution}
The finite number of detected particles in detectors produces a limited resolution in the measured event plane angle. So, the observed flow coefficients must be corrected up to what they would be relative to the real reaction plane. This is done by dividing these coefficients by the event plane resolution, estimated from the correlation of the planes of independent subevents. 
\begin{equation}
    v_{n} = \frac{v_n^{obs}}{R_n} = \frac{v_n^{obs}}{\langle\cos[n(
    \psi_n - \Psi_R)]\rangle},
\end{equation}
\noindent
where the $R_{n}$ is resolution, $v_n$ is the $n^{th}$ harmonic azimuthal anisotropy parameter, and $\psi_m$ is the $m^{th}$ harmonic order event plane, $\Psi_{r}$ is reaction plane angle. The angle brackets denote an average over all particles in all events \cite{Ref4}. \\
In fixed target mode, the final state particle's acceptance is not symmetric around midrapidity. Therefore, the commonly used 2-sub event method, employed in the collider BES-I analysis, cannot be used to calculate the resolution. This method necessitates each sub-event to have similar multiplicity and resolution. Consequently, in this analysis, we opt for the 3-sub event method to calculate the resolution. Figure 2 shows the calculated first-order event plane resolution $R_{11}$ and the third-order event plane resolution $R_{13}$ estimated from the first-order event plane for $v_3$ calculation, as functions of collision centrality.
\renewcommand{\thefigure}{2}
\begin{figure}
\begin{tabular}{ccc}
\includegraphics[width=4.8cm,height=5cm]{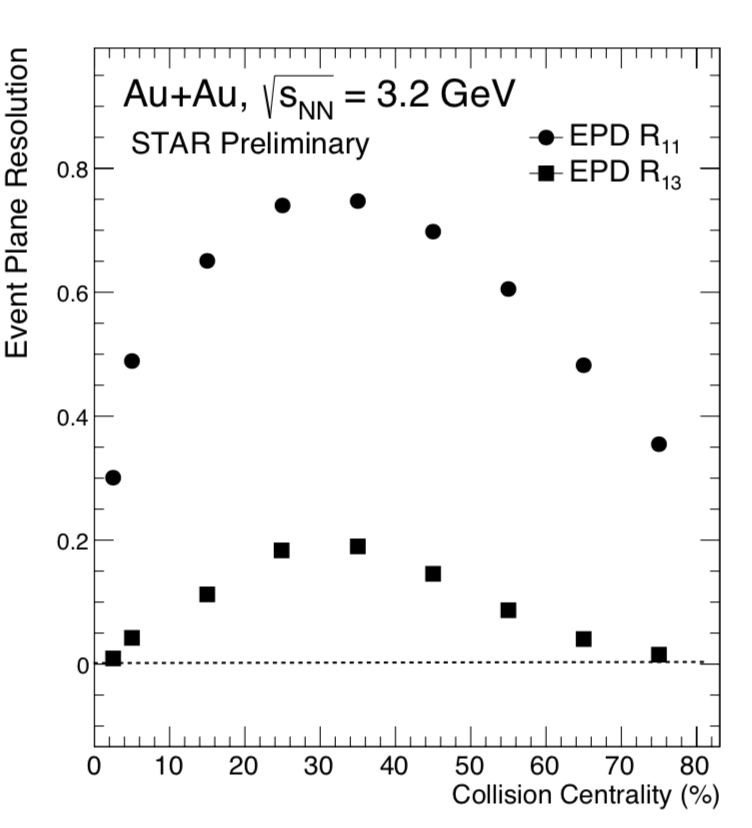}
\includegraphics[width=4.8cm,height=5cm]{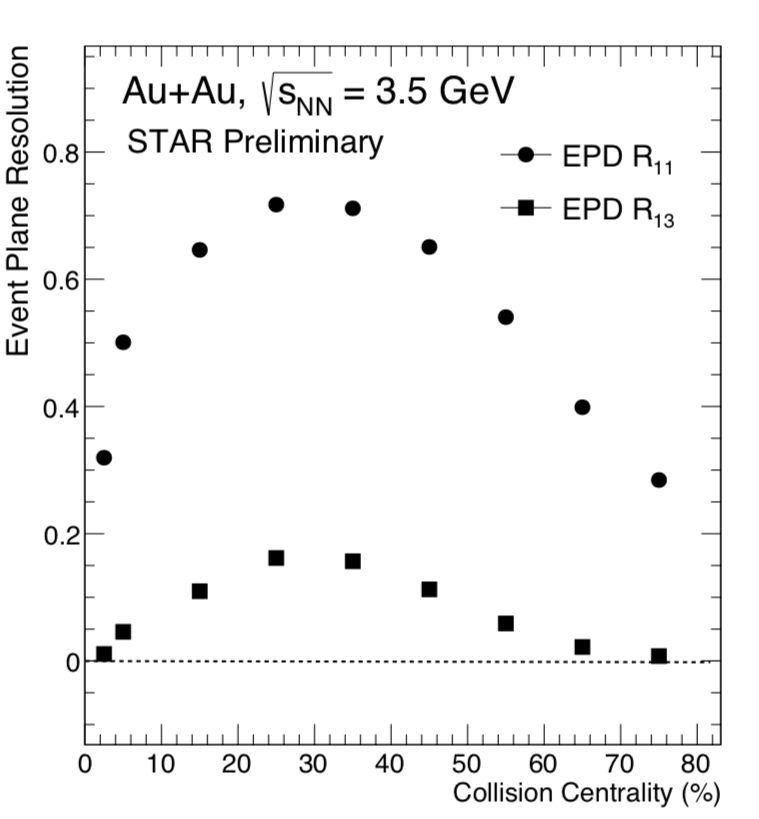}
\includegraphics[width=4.8cm,height=5cm]{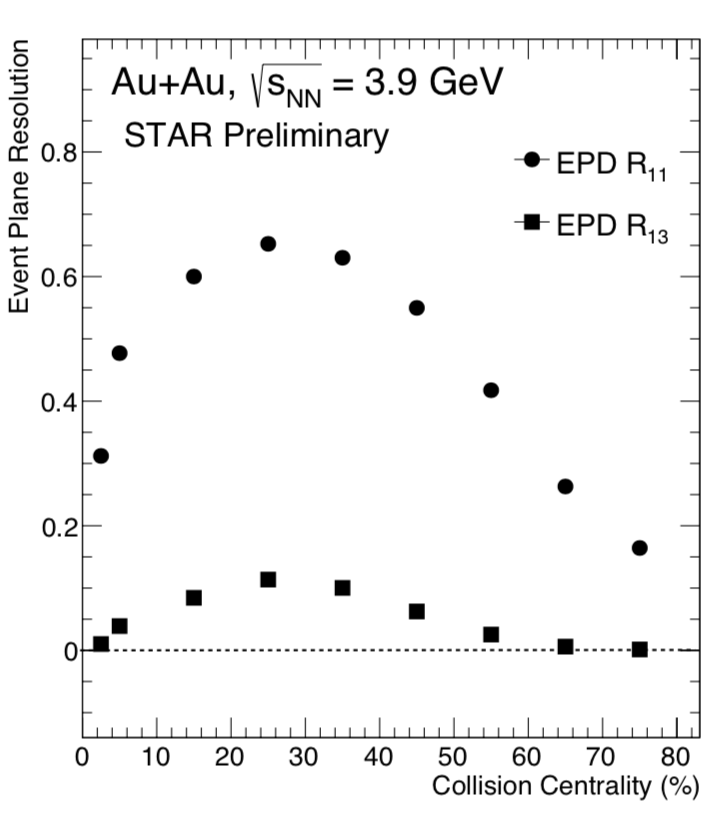}
\end{tabular}
\centering
\caption{Collision centrality dependence of $R_{11}$ (circles) $R_{13}$ (squares) in Au+Au collisions at $\sqrt{s_{\text{NN}}}$= 3.2 (left panel), 3.5 (middle panel), and 3.9 GeV(right panel).}
\end{figure}

\section{Results and Discussion}
\subsection{Directed Flow $(v_1)$}
The rapidity (y), centrality and collision energy dependence of $v_1$ for identified hadrons, net-particle, and light nuclei are measured at $\sqrt{s_{\text{NN}}}$= 3.2, 3.5, and 3.9 GeV. \\ 
\renewcommand{\thefigure}{3}
\begin{figure}
\begin{tabular}{c}
\includegraphics[width=14.0cm,height=5.0cm]{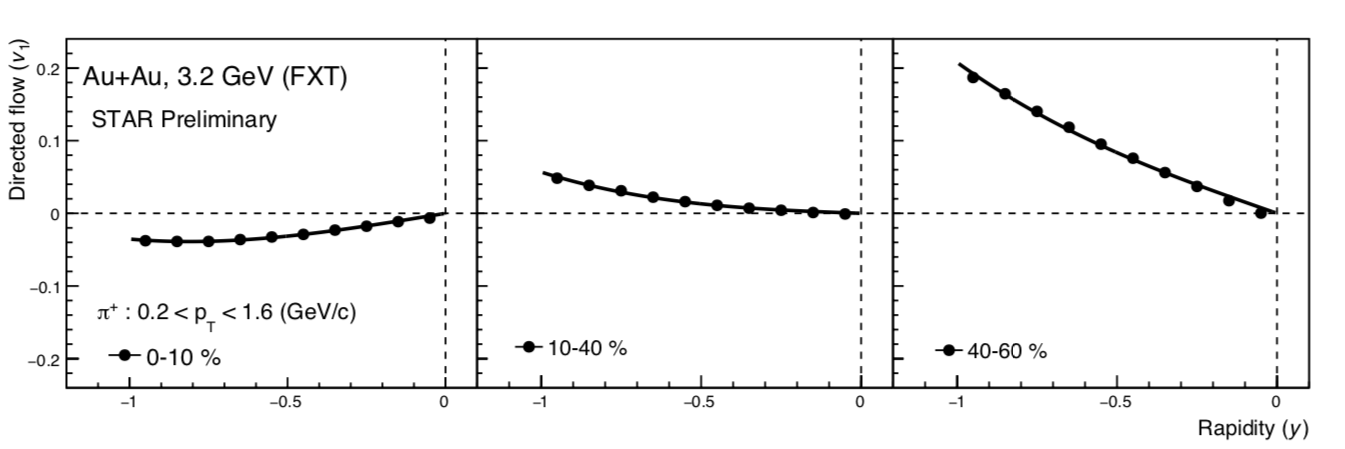}
\end{tabular}
\centering
\caption{$v_1$ as a function of y for pion in 0-10\% (left panel), 10-40 $\%$ (middle panel), and 40-60\% (right panel) centrality bin in Au+Au collisions at $\sqrt{s_{\text{NN}}}$ = 3.2 GeV. The line represents third order polynomial fit to distribution.}
\end{figure}
Figure 3 illustrates the the centrality dependence of $\pi^{+}$ for $\sqrt{s_{NN}}$= 3.2 GeV. The $v_{1}$ changes sign from negative to positive, moving from most central to peripheral collisions, implying the effect of dominant repulsive baryonic interactions and spectator shadowing.\\

\noindent The energy dependence of proton $v_1$ involves an interplay between the directed flow of protons associated with baryon stopping and particle-antiparticle pair production at mid-rapidity. A means to distinguish between the two mechanisms would thus be to look at the net particle $v_1$. The net particle represents the excess yield of a particle species over its antiparticle. The net particle's $v_{1}$ is defined as 
\begin{equation}
    v_{1, net} = \frac{v_{1,p} -rv_{1,\bar{p}}}{1-r},
\end{equation}

\noindent
where $v_{1,p}$, $v_{1,\bar{p}}$ corresponds to $v_{1}$ of particle and anti-particle, and $r$ represents the ratio of anti-particles to particles \cite{Ref9}. Figure 4 shows the y dependence of identified hadrons (left panel), net-particles (middle panel), and light nuclei (right panel) for $10-40\%$ centrality. The magnitude of $v_1$ increases with increasing rapidity for all particles, and a mass ordering is also observed in the magnitude of $v_1$.

\renewcommand{\thefigure}{4}
\begin{figure}
\begin{tabular}{ccc}
\includegraphics[width=4.8cm,height=4.8cm]{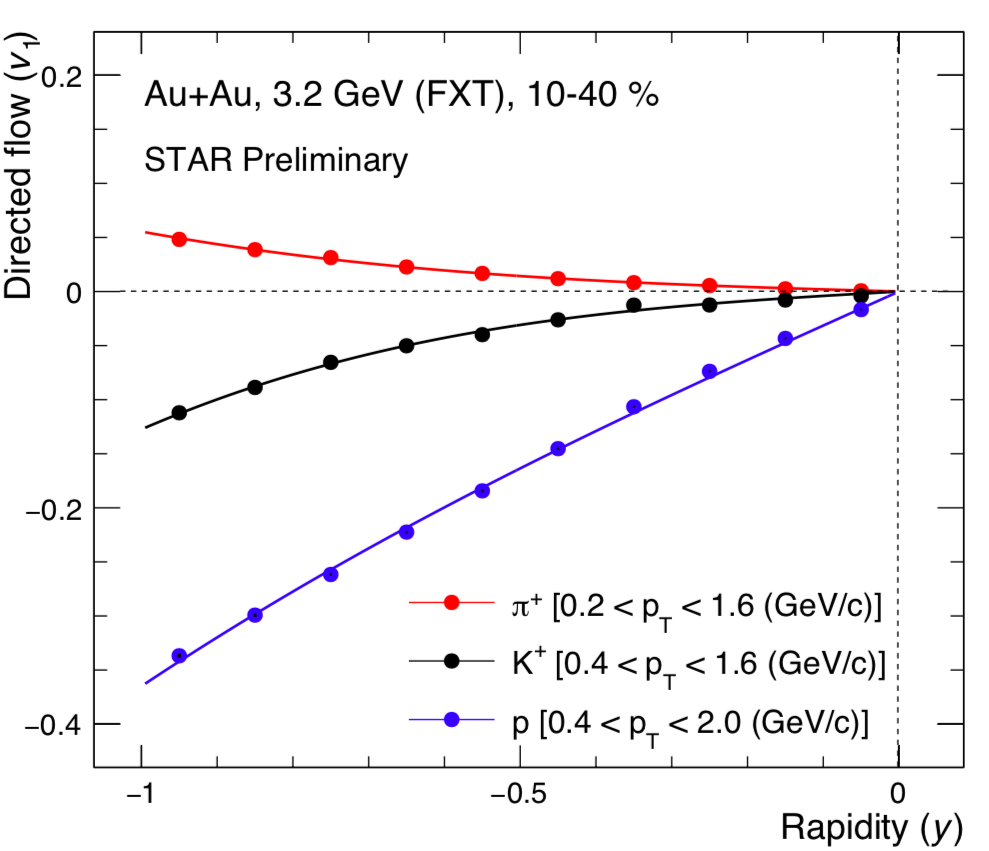}
\includegraphics[width=4.8cm,height=4.8cm]{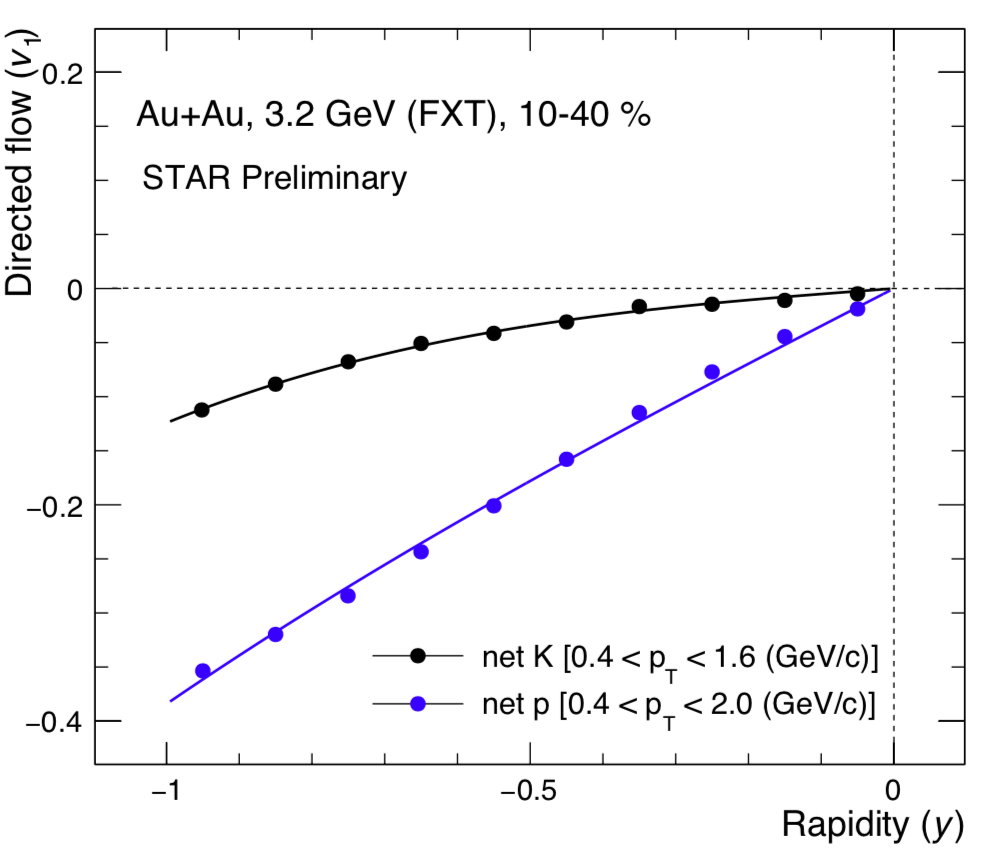}
\includegraphics[width=4.8cm,height=4.8cm]{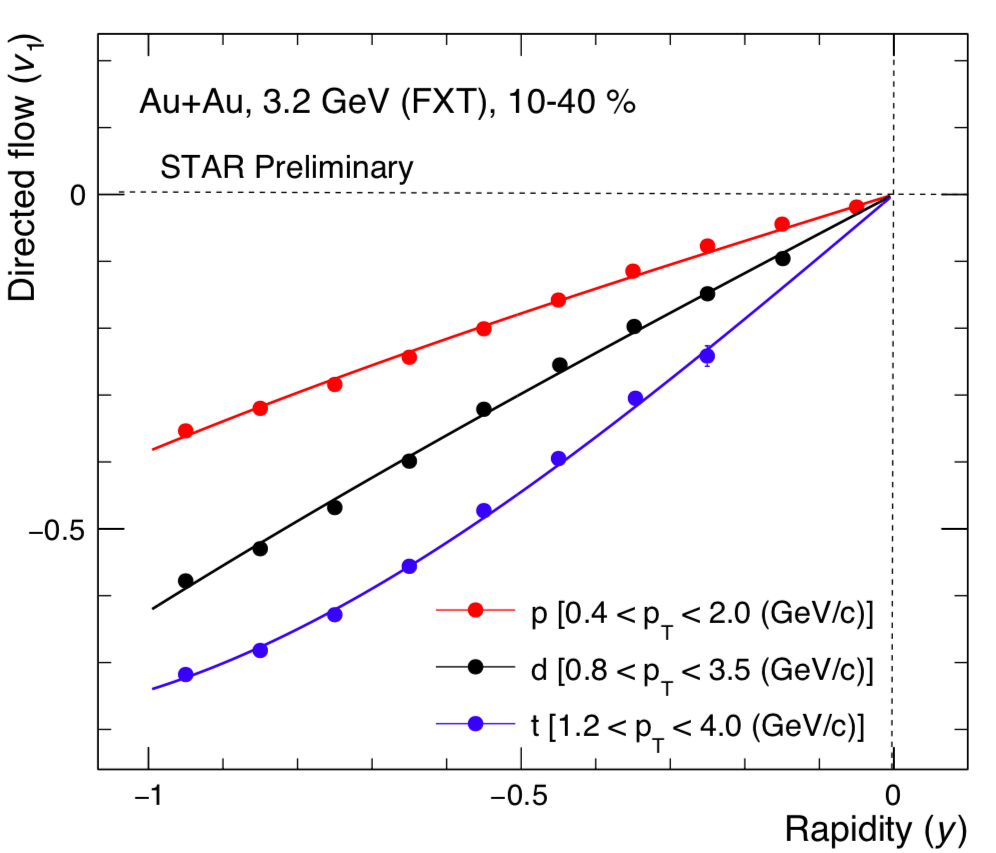}
\end{tabular}
\centering
\caption{$v_1$ as a function of y in $10-40\%$ centrality for identified hadrons (left panel), net particles (middle panel) and light nuclei (right panel) in Au+Au collisions at $\sqrt{s_{\text{NN}}}$ = 3.2 GeV. The line represents $3^{\text{rd}}$ order polynomial fit to distribution.}
\end{figure}
The $p_T$-integrated $v_1 (y)$ slope at mid-rapidity, $dv_1/dy|_{y=0}$, is obtained by fitting the data $v_1(y)$ with a third-order polynomial. Figure 5 shows the collision energy dependence of $dv_1/dy|_{y=0}$ for identified particles (left panel), net-particle (middle panel), and light nuclei (right panel) in mid-central (10 - 40\%) collisions.
The extracted slope parameters, $dv_1/dy|_{y=0}$, are scaled by A for light nuclei to compare with protons. The magnitude of the slope decreases with increasing collision energy for all particles, including light nuclei. The slope of $v_1$ for net-kaon undergoes a sign change from negative to positive at a lower collision energy range ($\sqrt{s_{\text{NN}}}$  = 3.9 - 7.7 GeV) compared to net-proton ($\sqrt{s_{\text{NN}}}$  =  11.5 - 19.6 GeV). The light nuclei $v_1$ slope exhibits an approximate mass number (A) scaling, consistent with the nucleon coalescence mechanism for the production of light nuclei. 

\renewcommand{\thefigure}{5}
\begin{figure}
\begin{tabular}{ccc}
\includegraphics[width=4.8cm,height=5.2cm]{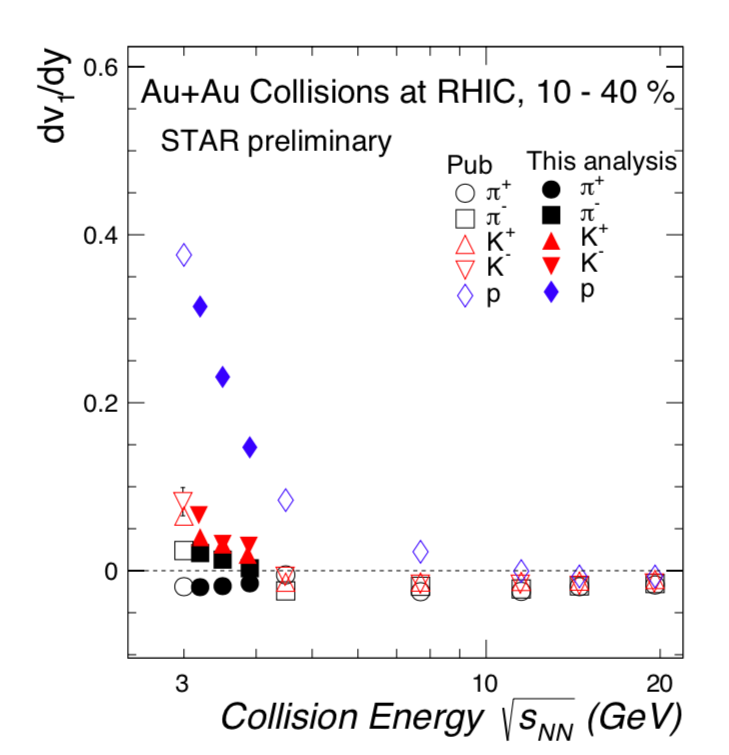}
\includegraphics[width=4.8cm,height=5.2cm]{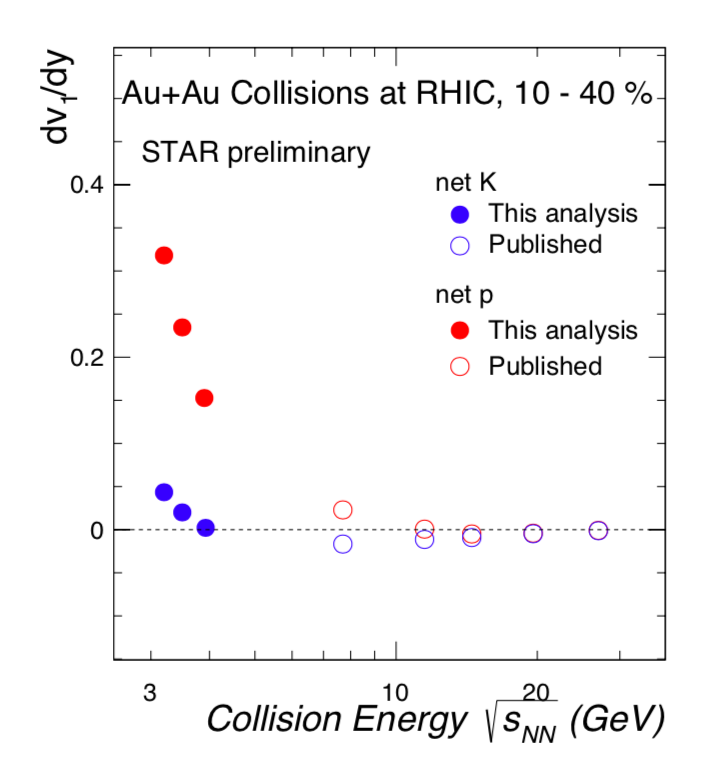}
\includegraphics[width=4.8cm,height=5.2cm]{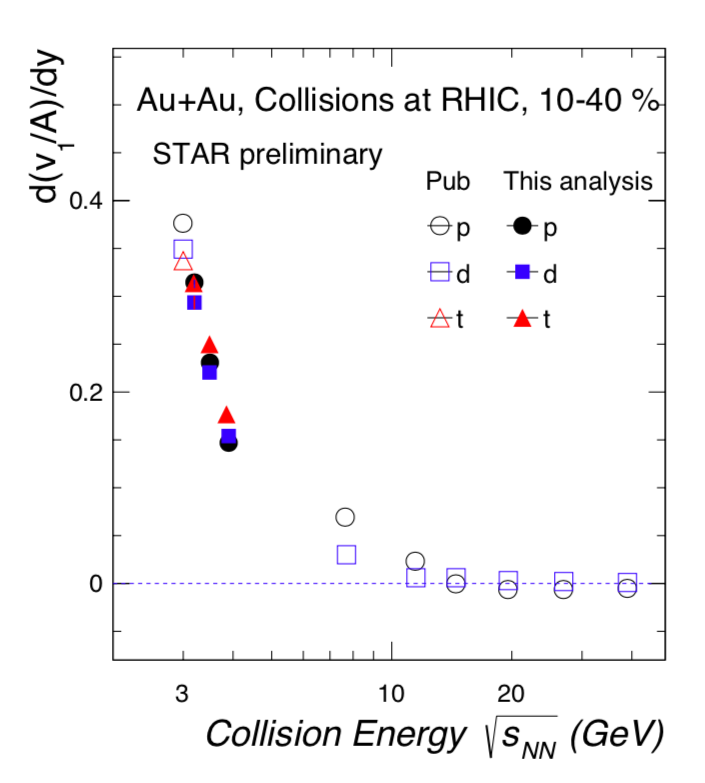}
\end{tabular}
\centering
\caption{Collision energy dependence of $v_1$ slope $\textrm{d}v_1/\textrm{d}y|_{y=0}$ for identified hadrons (left panel), net particles (middle panel) and light nuclei (right panel) in Au+Au collisions at RHIC for $10-40\%$ centrality. The published data are shown in open markers~\cite{Ref12}.}
\end{figure}

\subsection{Triangular Flow $(v_{3})$}
The y and collision energy dependence of $v_3$ for identified hadrons and light nuclei are measured at $\sqrt{s_{\text{NN}}}$= 3.2, 3.5, and 3.9 GeV. The left panel of Fig. 6 shows the rapidity dependence of $v_3$ for identified hadrons. The magnitude of $v_3$ increases with increasing rapidity. The distribution is fitted with a polynomial of order three to extract the slope parameter. \\
The middle panel of Fig. 6 shows the slope of $v_3$, $dv_3/dy|_{y=0}$, for identified hadrons as a function of collision energy. The magnitude of $dv_3/dy|_{y=0}$ decreases with increasing collision energy. It may indicates that the combined effect of the mean-field, baryon stopping, and collision geometry is considerably significant at the low collision energies~\cite{Ref13}.\\
The right panel of Fig. 6 shows the extracted slope parameters, $dv_3/dy|_{y=0}$, scaled by mass number (A) for light nuclei. The magnitude of the slope decreases with increasing collision energy. The light nuclei $v_3$ slope also exhibits an approximate mass number (A) scaling, consistent with the nucleon coalescence mechanism for the light nuclei production.

\renewcommand{\thefigure}{6}
\begin{figure}
\begin{tabular}{ccc}
\includegraphics[width=4.6cm,height=5.cm]{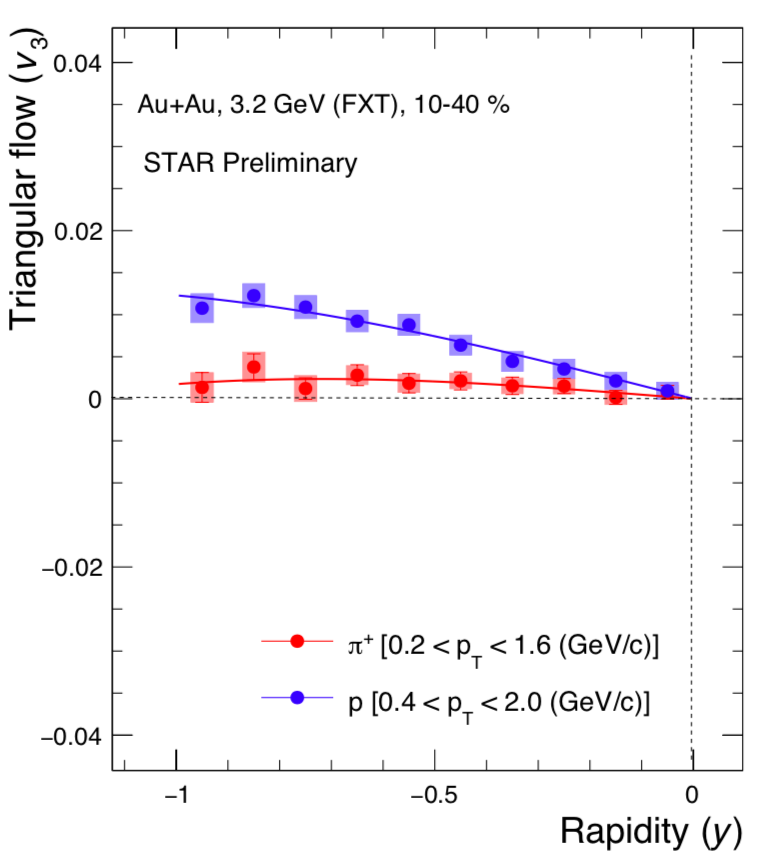}
\includegraphics[width=4.8cm,height=5.2cm]{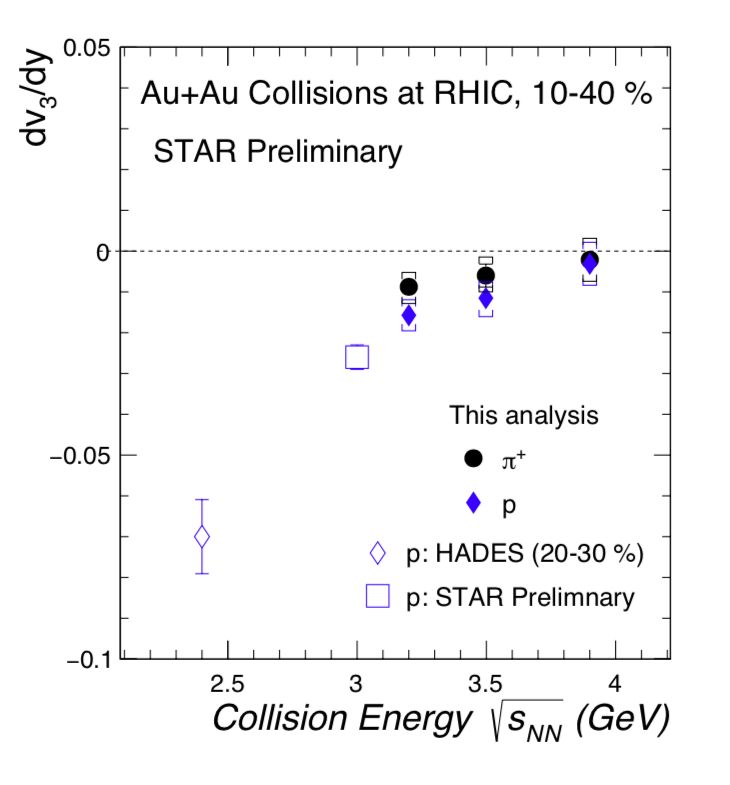}
\includegraphics[width=4.8cm,height=5.2cm]{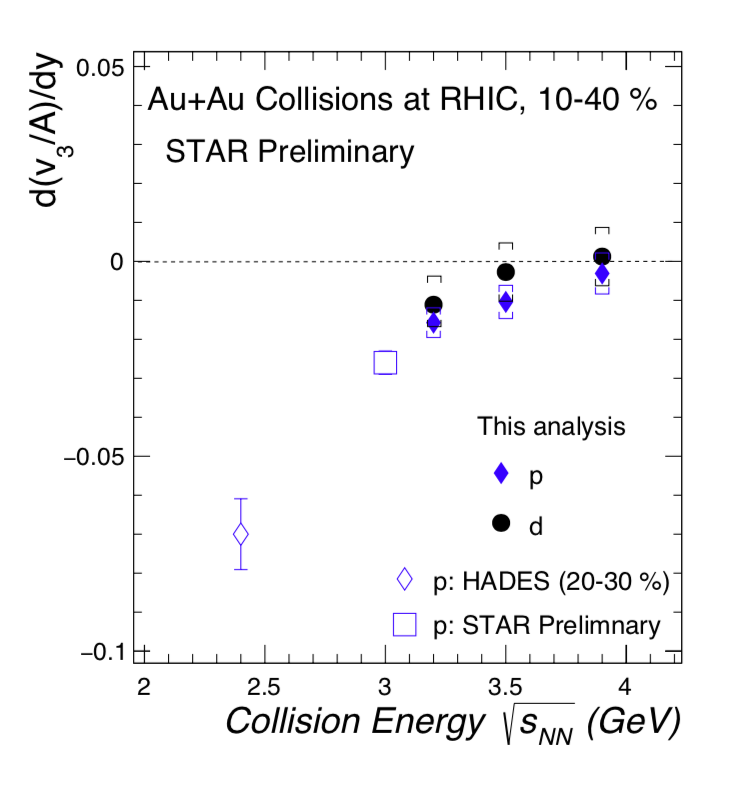}
\end{tabular}
\centering
\caption{$v_3$ as a function of y in $10-40\%$ centrality bin for identified hadrons (left panel) in Au+Au collisions at $\sqrt{s_{\text{NN}}}$ = 3.2 GeV. Collision energy of $\textrm{d}v_3/\textrm{d}y|_{y=0}$ for identified particles (middle panel) and light nuclei (right panel) in Au+Au collisions at RHIC for $10-40\%$ centrality. The published data are shown in open markers~\cite{Ref10}.}
\end{figure}

\section{Conclusion}
In summary, the rapidity, centrality, and collision energy dependence of directed flow $(v_1)$ of identified hadrons, net particle, and light nuclei in Au+Au collisions at $\sqrt{s_{NN}}$ = 3.2, 3.5, and 3.9 GeV is reported. The magnitude of $v_1$ increases with increasing rapidity for all particles. The extracted $v_1$ slope of all the particles decreases in magnitude with increasing collision energy. A positive $v_1$ slope at mid-rapidity for identified hadrons and net particles, excluding $\pi^{+}$, suggests prevalent repulsive baryonic interactions and spectator shadowing. As collision energy decreases, a non-monotonic trend is observed in the slope of both net-kaon and net-proton. The $v_1$ slope for net-kaon experiences a transition from negative to positive at a collision energy lower than that observed for net-proton.
The light nuclei $v_1$ slope exhibits an approximate mass number scaling consistent with the nucleon coalescence mechanism for the production of light nuclei. The magnitude of slope of $v_3$ decreases with increasing collision energy, indicating a substantial collective impact of the mean-field, baryon stopping, and collision geometry at lower collision energies.

\bibliographystyle{plain}

\end{document}